\documentclass{pasj00}
\begin{document}
\SetRunningHead{M. Kawada}{Imaging FTS for AKARI}
\Received{2008/06/00}%{yyyy/mm/dd}
\Accepted{2008/00/00}%{yyyy/mm/dd}

\title{Performance of the Imaging Fourier Transform Spectrometer 
with Photoconductive Detector Arrays:
\\ An Application for the AKARI Far-Infrared Instrument}

%%% begin:list of authors
% Do NOT capitalize all letters in "textsc".
 \author{%
   Mitsunobu \textsc{Kawada}\altaffilmark{1},
   Hidenori \textsc{Takahashi}\altaffilmark{2},
   Noriko \textsc{Murakami}\altaffilmark{1,3},
   Hiroshi \textsc{Matsuo}\altaffilmark{4},
   Yoko \textsc{Okada}\altaffilmark{5},
   Akiko \textsc{Yasuda}\altaffilmark{5},
   Shuji \textsc{Matsuura}\altaffilmark{5},
   Mai \textsc{Shirahata}\altaffilmark{5},
   Yasuo \textsc{Doi}\altaffilmark{6},
   Hidehiro \textsc{Kaneda}\altaffilmark{5},
   Takafumi \textsc{Ootsubo},\altaffilmark{5}
   Takao \textsc{Nakagawa}\altaffilmark{5},
   and
   Hiroshi \textsc{Shibai}\altaffilmark{1,7}}
\altaffiltext{1}{Graduate School of Sciences, Nagoya University,
	Furo-cho, Chikusa-ku, Nagoya 464-8602, Japan}
 \email{kawada@u.phys.nagoya-u.ac.jp}
\altaffiltext{2}{Gunma Astronomical Observatory, 6860-86 Nakayama, 
	Takayama, Agatsuma, Gunma 377-0702, Japan}
\altaffiltext{3}{Bisei Astronomical Observatory, Okura, Bisei-cho,
	Ibara-shi, Okayama 714-1411, Japan}
\altaffiltext{4}{Advanced Technology Center, 
	National Astronomical Observatory of Japan,
	2-21-1 Osawa, Mitaka, Tokyo 181-8588, Japan}
\altaffiltext{5}{Institute of Space and Astronautical Science, JAXA,
	3-1-1 Yoshinodai, Sagamihara 229-8510, Japan}
\altaffiltext{6}{Department of General System Studies, 
	Graduate School of Arts and Sciences, 
	The University of Tokyo, 3-8-1 Komaba, Meguro-ku, 
	Tokyo 153-8902, Japan}
\altaffiltext{7}{Graduate School of Sciences, Osaka University,
	1-1 Machikaneyama-cho,Toyonaka-shi,Osaka 560-0043,Japan}
%%% end:list of authors

%%% Please use the following style in case that sorting by 
%%% affilation is impossible. 
%
%% `\KeyWords{}' always has to be placed before `\maketitle'.
 %Do NOT move this preamble from here!

\maketitle

\begin{abstract}
We have developed an imaging Fourier transform spectrometer (FTS) for 
space-based far-infrared astronomical observations. The FTS employs a 
newly developed photoconductive detector arrays with a capacitive 
trans-impedance amplifier, which makes the FTS a completely unique 
instrument. The FTS was installed as a function of the far-infrared 
instrument (FIS: Far-Infrared Surveyor) on the Japanese astronomical 
satellite, AKARI, which was launched on February 21, 2006 (UT) from 
the Uchinoura Space Center. The FIS-FTS had been operated for more 
than one year before liquid helium ran out on August 26, 2007. The 
FIS-FTS was operated nearly six hundreds times, which corresponds to 
more than one hundred hours of astronomical observations and almost 
the same amount of time for calibrations. As expected from laboratory 
measurements, the FIS-FTS performed well and has produced a large 
set of astronomical data for valuable objects. Meanwhile, it becomes 
clear that the detector transient effect is a considerable factor 
for FTSs with photoconductive detectors. In this paper, the 
instrumentation of the FIS-FTS and interesting phenomena related 
to FTS using photoconductive detectors are described, and future 
applications of this kind of FTS system are discussed.
\end{abstract}

\section{Introduction}

Observations using infrared wavelengths are progressing in many fields 
of science and technology. In astrophysics, space-based infrared 
astronomy has become a powerful tool for deep insights into the universe 
ever since the first astronomical satellite, IRAS, was launched in 1983 
\citep{Neugebauer84}.   
The earliest instruments for infrared astronomy were photometers using 
discrete detectors, and observations become gradually more precise, 
driven by the evolution of infrared technology. The development of 
detector arrays allowed detailed images to be recorded efficiently, 
and when coupled with spectrometers, provided detailed imaging 
spectroscopy. The COBE satellite 
\citep{Mather93a}, 
which was launched in 1989, measured the precise spectra and absolute 
power of the cosmic infrared background radiation. 
IRTS \citep{Murakami96}, which was launched in 1995 as the first 
Japanese infrared telescope in space, have mapped 7 \% of the sky, 
and have provided near- \citep{Tanaka96}, mid- \citep{Onaka96}, and 
far- \citep{Shibai96} infrared spectra and line intensities as well as 
sub-millimeter brightness distribution \citep{Hirao96}.   
The ISO \citep{Kessler96, Kessler02}, 
which was launched in 1995, provided a lot of infrared spectra for 
various objects, and emphasized the importance of studying infrared 
lines for investigations of the interstellar medium. The Spitzer Space 
Telescope 
\citep{Werner04}, 
the infrared astronomical satellite launched in 2003, provides 
valuable images and spectra for various objects, but has only limited 
spectroscopic capabilities in the far infrared wavelength range.   

The AKARI satellite \citep{Murakami2007} 
is the Japanese infrared astronomical satellite, which was launched in 
2006. AKARI has a 68.5 cm diameter cooled telescope \citep{Kaneda2007} 
and two focal plane instruments; one is the IRC (InfraRed Camera) 
\citep{Onaka2007} 
for near to mid infrared range and the other is the FIS (Far-Infrared 
Surveyor) \citep{Kawada2007} 
for far-infrared imaging photometry. The FIS has spectroscopic 
capabilities as an optional function, provided by a Fourier transform 
spectrometer (FTS) \citep{Takahashi03}.  
The FTS of the FIS (hereafter FIS-FTS) is the second spectrometer for 
far-infrared astronomy operated in space after FIRAS \citep{Mather93b} 
of the COBE satellite. A unique feature of the FIS-FTS is the imaging 
FTS design using a photoconductive detector system. The design, 
operation and performance of the FIS-FTS are described in this paper; 
the data reduction and the calibration of the FIS-FTS are treated in 
the companion paper \citep{Murakami2008}, 
and some scientific results related to the FIS-FTS are discussed by 
some authors \citep{Yasuda2008, Okada2008, Takahashi2008}. 
In the next section, the optical and mechanical design of the FIS-FTS 
are described. After explaining the operation of the FIS-FTS in 
section 3, the performance and some interesting properties of the 
FIS-FTS are presented in section 4. In section 5, we discuss special 
topics related to the FIS-FTS and future applications of FTSs with 
photoconductive detectors.

\section{Instrumentation}

\subsection{Optical Design}

%\ref{fig:FIS_optics}
\begin{figure*}
  \begin{center}
    \FigureFile(150mm,100mm){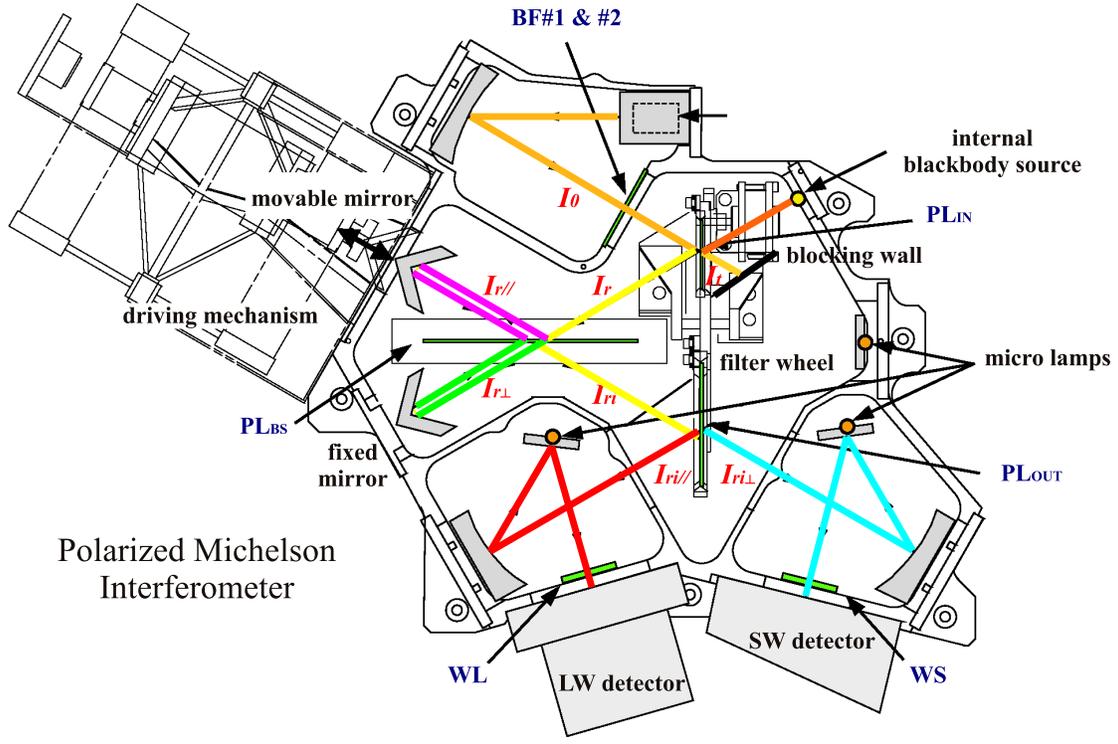}
  \end{center}
  \caption{Drawings of the FIS optical configuration. The major optical 
	components are shown with labels. The FIS instrument provides two 
	functions --- the photometer (right side) and the spectrometer (left 
	side), which use different optical paths.  Each function is 
	selected by rotating the filter wheel. }
  \label{fig:FIS_optics}
\end{figure*}

The FIS-FTS adopts a polarized Michelson-type Martin-Puplett 
interferometer (MPI) design \citep{Martin69}. 
MPIs require a linear polarized light for input, which is then divided 
into two polarization components by a polarizing beam splitter. 
The two components are interfered as an elliptically polarized beam 
according to the incident spectrum and the optical path difference 
between the two components. By picking up the polarization component 
parallel or perpendicular to the polarization angle of the incident 
beam, interferograms can be recorded.

A schematic view of the FIS-FTS optics is shown in figure  \ref{fig:FIS_optics}. 
The photometer and the FTS are individually selected by rotating the 
filter wheel. The collimator and camera optics, some optical filters 
and the detector system are commonly used in both functions. The 
incident beam ($I_0$) is collimated by the collimator mirror and passed 
to the input polarizer (PL$_{IN}$) on the filter wheel through the blocking 
filters (BF\#1 \& \#2).  
The input polarizer (PL$_{IN}$) produces linear polarization; the reflected 
component ($I_r$) 
goes into the interferometer and the transmitted component ($I_t$) 
is absorbed by a black wall. The linear polarized beam ($I_r$) 
is divided by the polarizing beam splitter (PL$_{BS}$), whose polarization angle 
is rotated by 45 degrees to the polarization angle of the incident beam.  
The two divided components ($I_{r\parallel}$ and $I_{r\perp}$) 
are reflected by roof-top mirrors; one is fixed and the other is 
movable to produce an optical path difference. Since the polarization 
angle of each component is rotated by 90 degrees by roof-top mirrors, the 
reflected component ($I_{r\parallel}$) 
is transmitted and the transmitted component ($I_{r\perp}$) 
is reflected by the beam splitter (PL$_{BS}$). The two interfering beams ($I_{ri}$) 
are modulated as an elliptical polarization and divided into two axis 
components by the output polarizer (PL$_{OUT}$) on the filter wheel, whose polarizing 
angle is the same with that of the input polarizer. The component reflected 
by the output polarizer ($I_{ri\parallel}$) 
is focused on the LW detector through the low pass filter (WL) which covers 
the 55 -- 90 cm$^{-1}$ wavenumber range (110 -- 180 $\micron$ in wavelength), 
and the other component ($I_{ri\perp}$) is focused on the SW detector through 
the low pass filter (WS) which covers the 90 -- 200 
cm$^{-1}$ wavenumber range (50 -- 110 $\micron$ in wavelength). 
The modulated powers on each detector ($I_{ri\parallel}$ and $I_{ri\perp}$) 
are described for incident photons of a wavenumber $\sigma$,
\begin{equation}
  I_{ri\parallel}(\sigma) = \frac{I_r}{2} (1 + cos 2 \pi \sigma x),  \nonumber \\
  I_{ri\perp}(\sigma) =\frac{I_r}{2} (1 - cos 2 \pi \sigma x) 
  \label{eq:interferogram}
\end{equation}
where the $x$ 
is the optical path difference. As shown in eq. (\ref{eq:interferogram}), 
the polarities of the modulated signals are opposite each other.

To reduce the size of the optics, the incident angles of the polarizers 
are 30 and 60 degrees instead of the usual 45 degrees. All polarizers 
are wire grid type filters. The structure of the 
filters is thin copper wires printed on thin Mylar film mounted in a 
stainless frame.  All optical filters used in the FIS-FTS are listed 
in table \ref{tbl:filters}, which were supplied by QMC Ins. Ltd.
As the result of the optical configuration of MPIs, only one fourth of 
the incident power is utilized for each detector for unpolarized 
incident flux, even in the ideal case. The actual optical efficiency is 
degraded by the quality of the optical components. Furthermore, the 
modulation efficiency depends on the optical alignment of the 
interferometer and on the detector properties.  The performance of the 
FIS-FTS is described in section \ref{ss:performance}.
 
\begin{table*}
  \caption{Specifications of optical filters using in the FIS-FTS.}\label{tbl:filters}
  \begin{center}
    \begin{tabular}{lllll}
      \hline
	GROUP & NAME & TYPE & EFFECTIVE SIZE & PROPERTY \\
	\hline
      blocking filter & BF\#1 & multi mesh LPF & 27 mm$\phi$ & $\sigma_{cut}$ = 430 cm$^{-1}$ \\
	& BF\#2 & multi mesh LPF & 27 mm$\phi$ & $\sigma_{cut}$ = 300 cm$^{-1}$ \\
      polarizing filter & PL$_{IN}$ & wire grid & 30 mm$\phi$ &  \\
	 & PL$_{BS}$ & wire grid & 85 mm$\phi$ &  \\
	 & PL$_{OUT}$ & wire grid & 56 mm$\phi$ &  \\
      shaping filter & WS & multi mesh LPF & 22 mm$\phi$ & $\sigma_{cut}$ = 218 cm$^{-1}$ (for SW) \\
	 & WL & multi mesh LPF & 22 mm$\phi$ & $\sigma_{cut}$ = 90 cm$^{-1}$ (for LW) \\
     \hline
	\multicolumn{5}{c}{ } \\
	\multicolumn{5}{l}{\hbox to 0pt{\parbox{160mm}{\footnotesize
		Notes: All filters were supplied by QMC Instruments Ltd.
	}\hss}}
    \end{tabular}
  \end{center}
\end{table*}

The FIS-FTS has two internal calibration sources; one consists of 
micro-lamps to monitor the detector responsivity and the other is a 
small blackbody source, whose temperature is controllable up to about 
50K. The blackbody source is put at the input of the interferometer 
opposite to the input polarizer.  The attenuation factor of the small 
blackbody source relative to the sky signal is about $10^{-3}$ 
due to its small aperture size of 0.7 mm$\phi$.  
As the results of the small aperture size of the internal blackbody 
source, the effects of vignetting for pixels at the edge of the 
detector arrays and the modulation efficiency due to the misalignment 
of the two beams are smaller than for sky signals. 
Furthermore, the polarity of the blackbody interferograms is opposite 
to those for astronomical sources, since the transmitted beam is used 
for the internal blackbody source.

\subsection{Mechanical Design}

\begin{figure}
  \begin{center}
    \FigureFile(80mm,60mm){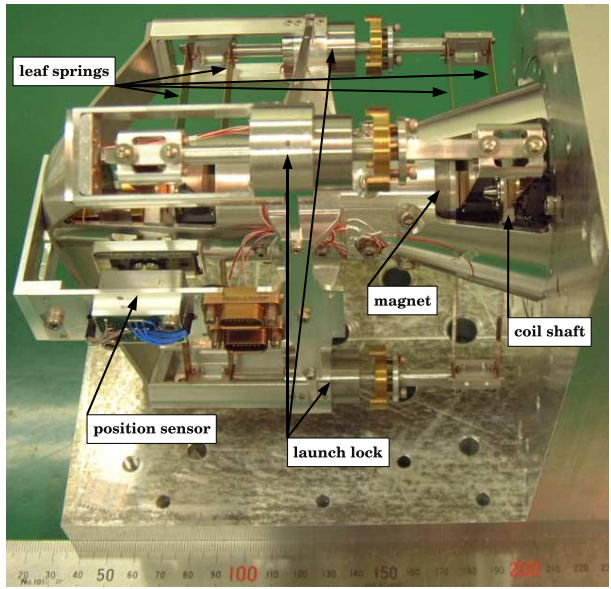}
  \end{center}
  \caption{A photograph of the mirror driving mechanism before installing on the FIS 
	instrument. Optical shield is removed to see the inside of the mechanism.}
  \label{fig:mirror_drive}
\end{figure}

To realize the interferometer described above, the mechanical and 
thermal designs were optimized for the operation in space and 
cryogenic environment. The fundamental mechanical component of the 
interferometer is the mirror driving mechanism which produces the 
optical path difference between the two interfering beams. The 
driving mechanism for the FIS-FTS is based on the mechanism of the 
CIRS instrument on the Cassini spacecraft \citep{Kunde96}.  
The roof-top mirror is mounted on a moving coil supported by four 
parallel tri-symmetrical leaf springs made of phosphor bronze, 
although the CIRS adopts a moving magnet. The coil, made of 
superconducting wire to reduce power dissipation, is enclosed in a 
magnet made of rare earth metals, which is designed to optimize both 
of the magnetic field and weight. The driving mechanism of the FIS-FTS 
is shown in figure  \ref{fig:mirror_drive}.  
The traveling length of the mirror driver is about $\pm$ 9.2 mm 
from the mechanical neutral position. The optical zero is set at the 
offset position about the half of the near side span. As the result of 
this configuration, the maximum optical path difference is about 
27.6 mm (twice of 13.8 mm) and the shorter side is about 9.2 mm.

%\ref{fig:mirror_drive}

To measure the mirror position, a Heidenhain type optical scale is 
adopted. The optical scale and sensor optics used are from a 
commercial grade Heidenhain LIP-401P module. The sensor unit, 
however, is modified because the original module does not work at 
cryogenic temperatures. The resolution of the mirror position is 1 $\micron$ 
at ambient temperature (half of the scale pitch), corresponding to 2 $\micron$ 
in optical path difference. However, the actual step size is reduced 
at the operation temperature of about 3K, because of its thermal 
contraction. Since we did not calibrate the scale in laboratory, we 
have to calibrate the scale in space using well known astronomical 
lines. Unfortunately, we have no sensor to detect the optical zero 
position. The optical zero position must be determined from the 
interferograms themselves. The specifications of the mirror 
mechanism is summarized in table \ref{tbl:FIS-FTS} 
with general properties of the FIS-FTS.  

The mirror driving mechanism is quite fragile, because of its 
requirements of optical accuracy and high mobility to reduce power 
consumption. This is not a problem for operation in space, but is 
troublesome as a result of the strong vibrations experienced during 
the launch. To survive the launch, a launch lock mechanism was adopted 
which fixes all movable parts by electromagnetic force during launch.  
The launch lock mechanism worked very well and the interferometer works 
as well in space as it did in the laboratory. 

\begin{table*}
  \caption{Specifications of the FIS-FTS.}\label{tbl:FIS-FTS}
  \begin{center}
    \begin{tabular}{lccl}
	\hline
	\multicolumn{4}{c}{Imaging Fourier Transform Spectrometer for Far-Infrared Surceyor} \\
	\hline
	Optics & \multicolumn{2}{c}{Polarized Michelson Interferometer} & Martin-Puplett type \\
	Mirror driving mechanism & \multicolumn{2}{c}{leaf springs axis support with voice coil actuator} & \\
	Position Sensor & \multicolumn{2}{c}{optical scale (2 $\mu$m in o.p.d.\footnotemark[$*$])} & {no zero path detector} \\
	Spectral coverage & \multicolumn{2}{c}{55 -- 200 cm$^{-1}$} & 50 -- 180 $\mu$m \\
	Maximum o.p.d.\footnotemark[$*$] ($L$) & \multicolumn{2}{c}{-0.9 -- +2.8 cm} & \\
	Spectra resolution (1/2$L$) & 0.19 cm$^{-1}$ (full res. mode) & 1.2 cm$^{-1}$ (SED mode) & without apodization \\
	Mirror scan speed & \multicolumn{2}{c}{$\sim$0.08 cm s$^{-1}$} & \\
	Detector system & \multicolumn{2}{c}{2D photoconductive detector arrays with CTIAs\footnotemark[$\dagger$]} & \\
	Detector unit name & SW & LW & \\
	Detector type & monolithic Ge:Ga array & stressed Ge:Ga array & \\
	Array format & 3 $\times$ 20 & 3 $\times$ 15 & \\
	Pixel size & 0.5 (0.55) mm & 0.9 (1.0) mm & (pixel pitch) \\
	Pixel FOV & 26.8$\times$26.8 arcsec$^2$ & 44.2$\times$44.2 arcsec$^2$ & \\
	PSF\footnotemark[$\ddagger$] (model) & \multicolumn{3}{l}{$P_{FTS}(x,y) = P_{phot}(x-7",y) + P_{phot}(x+7",y)$
		; ($x$: major axis of array)} \\
	 \quad FWHM & 44" / 39" & 57" / 53" & (major/minor) \\
	Spectral coverage & 85 -- 200 cm$^{-1}$ & 55 - 90 cm$^{-1}$ & \\
	Sampling rate & 85.336Hz & 170.672Hz & \\
	\hline
	\multicolumn{4}{c}{ } \\
	\multicolumn{4}{l}{\hbox to 0pt{\parbox{160mm}{\footnotesize
		Notes.
		\par\noindent
		\footnotemark[$*$] optical path difference
		\par\noindent
		\footnotemark[$\dagger$] Charge Trans-Impedance Amplifiers
		\par\noindent
		\footnotemark[$\ddagger$] Point Spread Function for the photometer: $P_{phot} = \varepsilon e^{-(x^2+y^2)/2{\sigma_1}^2}+(1-\varepsilon) e^{-(x^2+y^2)/2{\sigma_2}^2}$; \\ 
		 	$\varepsilon$ = 0.8; $\sigma_1$ = 14.2"; $\sigma_2$ = 40.0" (for SW), 
			$\quad \varepsilon$ = 0.65; $\sigma_1$ = 17.0"; $\sigma_2$ = 46.5" (for LW)
	}\hss}}
    \end{tabular}
  \end{center}
\end{table*}

\subsection{Detector System}
\label{sss:detector}

The FIS-FTS is an imaging FTS incorporating two, two-dimensional 
detector arrays. The detectors for the FIS-FTS are common with the 
photometer of the FIS instrument. There are two sets of detector array, 
one is a monolithic Ge:Ga array which covers 90 -- 200 cm$^{-1}$ 
(50 -- 110 $\micron$) (SW) 
\citep{Fujiwara03, Shirahata04} 
and the other is a compact stressed Ge:Ga array for 55 -- 90 cm$^{-1}$ 
(110 -- 180 $\micron$) (LW) \citep{Doi02}.  
The readout method of both detectors is a Charge Trans-Impedance 
Amplifier (CTIA) realized by cryogenic Silicon MOS-FETs 
\citep{Nagata04}.   
The CTIA integrates the photo-current of detector and is must be 
discharged periodically. The actual detector system has 
non-linearity due to the CTIA circuit itself, and additionally, the 
responsivity changes due to the drift of the effective detector bias 
voltage during the charge integration. Furthermore, extrinsic 
photoconductive detectors have a transient response with properties 
depending on both the background flux and the modulation amplitude 
\citep{Kaneda02}.  
These properties of the detector system make it difficult to analyze 
the interferometer output. In particular, since FTSs measure the 
modulation of the signal in the time domain, understanding the detector 
transient response is critical. Effects of these properties on the 
interferograms and spectra are described later.

\section{Operation}

The FIS-FTS is operated in time domain. The moving mirror is controlled 
without mirror position feedback. That is, the mirror position is 
controlled by certain fixed pattern of the driving current. The driving 
current pattern is optimized to drive the mirror with constant speed.  
Sampling timing is also determined in the time domain and not in the 
optical phase domain. As a result, the sampling interval is not 
constant in optical path difference, but constant in time. 
The nominal mirror speed is nearly 0.08 cm sec$^{-1}$ 
in optical path difference. The actual mirror speed has a global trend 
which is slower at the travel limits; the variation of the speed is 
less than $\pm$ 5\% 
across the whole travel span. The amplitude of the mirror speed 
fluctuation is also less than $\pm$ 5\% 
at the Nyquist sampling frequency without any disturbance. The normal 
sampling rate is nearly five times higher than the highest Nyquist 
frequency, which is helpful for reconstructing the interferograms from 
the integrated signal. After installing the FIS-FTS into the AKARI 
satellite, a mechanical interference between the mechanical cooler 
\citep{Nakagawa2007} 
and the mirror driving mechanism was recognized. The driving speed is 
modulated by the 15 Hz drive frequency of the cooler, with an amplitude 
of about $\pm$ 25\% 
of the mirror speed, reduced to about $\pm$ 15\% 
in space. Since the interference frequency is almost out of the optical 
modulation frequency corresponding to the effective wavelength ($<$ 15Hz), 
its influence can not be seen in spectra.
The sampling timing and the mirror driving pattern are synchronized by 
the FIS-FTS controller. All pixel signals for each detector array are 
acquired at nearly the same time as the mirror position. 
The data reduction method of the FIS-FTS is described by \citet{Murakami2008}.

For observation using the FIS-FTS, an astronomical observation template (AOT) 
is provided (see \citet{Kawada2007}), 
in which the calibration and the observation sequence are combined with 
some selectable parameters. Two driving current patterns are prepared 
for observations; one supports full travel span for high spectral 
resolution (full-res. mode) and the other reduces the span to one fourth 
but enables four times more scans (low-res. mode, which is called SED 
(Spectral Energy Distribution) mode).  
Driving speeds are almost the same in both modes.

\section{Performances}
\label{ss:performance}

\subsection{Optical Performances}
\subsubsection{Modulation Efficiency}
\label{sss:modulation_efficiency}
Optical performance of the interferometer is indicated by the modulation 
efficiency, which is defined as a ratio of the peak value of interferograms 
at the optical zero position to the constant signal level (constant part in 
eq. (\ref{eq:interferogram})).  The modulation efficiency is primarily 
affected by two factors; filter properties and optical alignment.

\begin{figure}
  \begin{center}
    \FigureFile(80mm,60mm){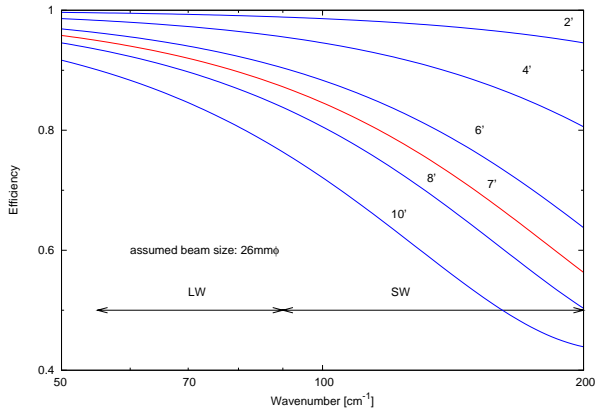}
  \end{center}
  \caption{Optical modulation efficiency expected as a function of misalignment of 
	the two beams. Several cases of misalignment are calculated and the seven 
	arcmin (red line) is the plausible case which is derived from the analysis 
	of zero path positions (see text for details).}
  \label{fig:mod_efficiency}
\end{figure}

Three polarizing filters are used in the FIS-FTS; all the same type of 
wire grid filter. The polarizing efficiency was measured for each in the 
laboratory and is more than 90\% 
for all. The resulting modulation efficiency due to the three polarizing 
filters in the MPIs configuration is about 80\%.  
The misalignment of the polarization angle of three filters and ridge 
angle of roof-top mirrors reduce this efficiency.  Since the 
misalignment of these angles was expected to be on the order of one 
degree, the effect of the misalignment to the modulation efficiency is 
negligible (less than 1\%).   
A major factor affecting the modulation efficiency is the misalignment 
of optical axes of the two beams, which is generated by the roof-top 
mirrors. That is, the opening angle of the roof-top mirrors and the 
alignment of the tilt angle to the optical axis. The opening angle of 
the roof-top mirrors is less than one arcmin from 90 degrees, which was 
measured in the laboratory at cryogenic temperature. The remaining 
misalignment of two beams is due to the tilt of the roof-top mirrors.  
The misalignment of the two beam axes can be derived from the 
interferograms. If the tilting angles of the two roof-top mirrors are 
the same, the optical zero will appear at the same time for all pixels.  
The actual optical zero for each pixel, however, appear at different 
times; there is a delay along the major axis of the detector array. 
From the measurement of this delay, the displacement angle of the two 
roof-top mirrors could be estimated at about 3.5 arcmin in orbit, which 
corresponds to a misalignment of about 7 arcmin for the two beams. This 
value is consistent with the laboratory measurement, and means that 
there is no degradation in optical alignment due to the launch. The 
expected optical modulation efficiency was calculated for several 
degrees of misalignment. As shown in figure  \ref{fig:mod_efficiency}, 
the optical modulation efficiency due to the misalignment of two beams 
is about 95--85\% for LW detector and about 85--60\% for SW detector.

The modulation efficiency can be estimated from the measured signal ratio 
of the peak value of the interferogram at the optical zero to the DC value, 
which indicates the total incident power containing both coherent and 
incoherent components. Due to the transient response of the detectors, 
however, it is difficult to estimate the actual modulation efficiency 
from the interferograms. In suction 5, we discuss the interferometric 
efficiency of the FIS-FTS estimated from the derived spectra, which 
include the detector transient response.

%\ref{fig:mod_efficiency}

\subsubsection{Image Quality}

%\ref{fig:flat}
\begin{figure}
  \begin{center}
    \FigureFile(80mm,60mm){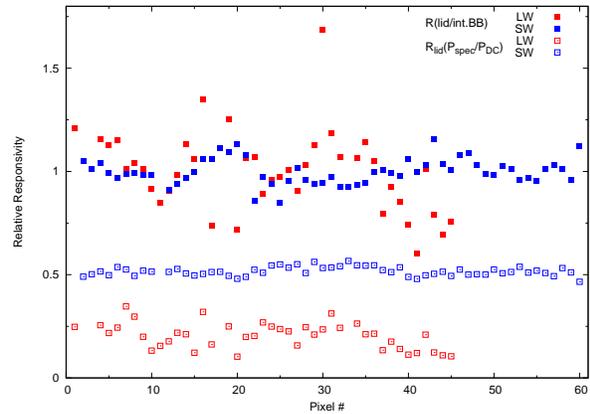}
  \end{center}
  \caption{The relative responsivity measured by telescope lid, which should be a flat source with 
	temperature of 35K.   The plot show the ratio of the integrated power of derived spectra of 
	the lid to that of the internal blackbody source (T=52K) or to that of the DC component.  
	The ratio of the lid to the internal blackbody (filled square) indicate the no uniformity 
	of the internal blackbody source primarily, though the ratio affected by the spectral 
	response and the transient response of the detector. The ratios to the DC component (open 
	square) indicates the interferometric efficiency, which includes the optical modulation 
	efficiency and the detector transient effect.  The plot of this value show wide variety 
	from pixel to pixel, especially for the LW detector.}
  \label{fig:flat}
\end{figure}

Image quality, which is a prominent feature of the FIS-FTS, was 
evaluated in orbit by observing Neptune with the FTS optics. The image of 
Neptune, which can be considered as a point source for the FIS 
instrument, was taken by a scanning observation mode. During the 
scanning observation, the movable mirror is in free motion away from the 
optical zero position ($\sim$ 0.9 cm 
in optical path difference). Neptune was observed with the same scanning 
mode with the photometer optics. The point spread functions (PSFs) of the 
photometer are well described by double Gaussian profiles, whose 
parameters are summarized in table \ref{tbl:FIS-FTS}.   
The resulting PSF of the FIS-FTS is rather broader than that of the 
photometer. The width of the PSF along the major axis of the detector 
array is wider, but there is no significant difference along the minor 
axis. This means that there is little affect on the image quality due 
to the FTS optics, i.e. the reflections at the surface of thin film 
polarizers. On the other hand, due to the misalignment of the roof-top 
mirrors, the image is elongated along the major axis of the detector.  
The elongated image of Neptune along the major axis of detector is 
consistent with an image derived from a beam misaligned by 7 arcmin, 
as described above.  The imaging spectroscopic performance is 
demonstrated by observations of the Galactic center region 
\citep{Yasuda2008}, 
the star forming regions on the Galactic plane \citep{Okada2008}, 
and nearby galaxies \citep{Takahashi2008}.

The FIS-FTS measures a wide wavenumber region (55 - 200 cm$^{-1}$) 
at the same time with two kinds of detector array. To produce the whole 
spectrum of a sky position, we have to know the alignment of the field 
of views (FOVs) of two detector arrays. The alignment of two arrays was 
determined from scanning observations of point like sources. The FOVs 
of two detector arrays are not aligned well on the sky as shown in 
figure  3 in \citet{Kawada2007}. 
The overlap region of the FOVs for the FIS-FTS is different from that 
of the FIS photometer, possibly arising from the alignment of the 
filter wheel. The smaller overlap region made it difficult to observe 
point like objects with both the arrays simultaneously.

Another factor affecting the image quality is a pixel-to-pixel 
variation in detector responsivity. During ground measurements of the 
instrument, the aperture lid of the cryostat was observed with the FIS-FTS as 
a flat source. The variations of the integrated signal power were 
about $\pm$ 30\% 
for the SW detector and $\pm$ 50\% 
for the LW detector, excluding a few extremely high pixels (see 
figure  \ref{fig:flat}). 
Since pixels of the LW detector were made from discrete chips, in 
contrast to the monolithic array of the SW detector, it exhibits much 
larger variations in the integrated signal power, which is coupled 
with the variation of the spectral response of each pixel. In 
combination with the internal blackbody source, which is not a flat 
source due to the influence of the FIS-FTS optics, we expect to 
improve the flatness of each detector to about $\pm$ 15\% 
and $\pm$ 30\% 
for the SW and LW detectors, respectively.

\subsection{Performance as a Spectrometer}

\subsubsection{Distortion of Interferograms}

\begin{figure*}
  \begin{center}
    \begin{tabular}{cc}
	\FigureFile(80mm,50mm){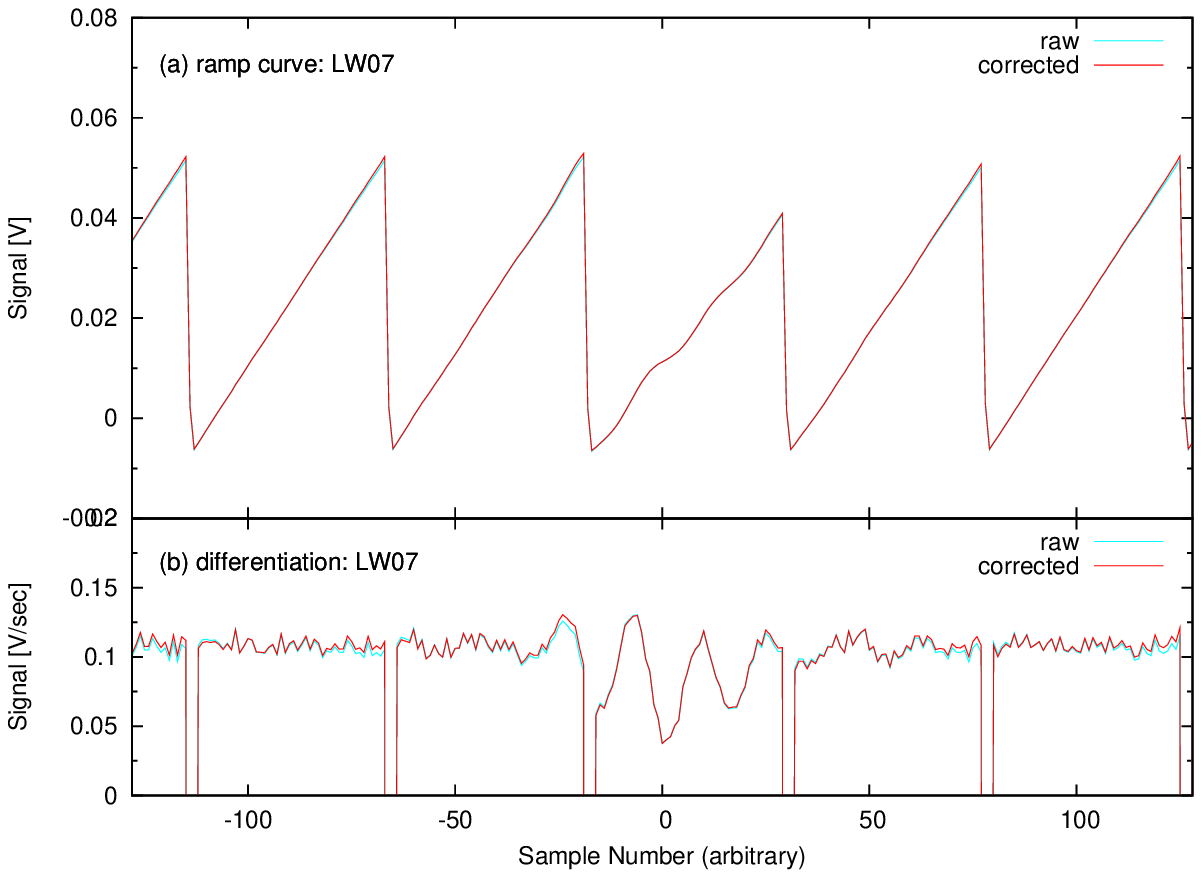} & \FigureFile(80mm,50mm){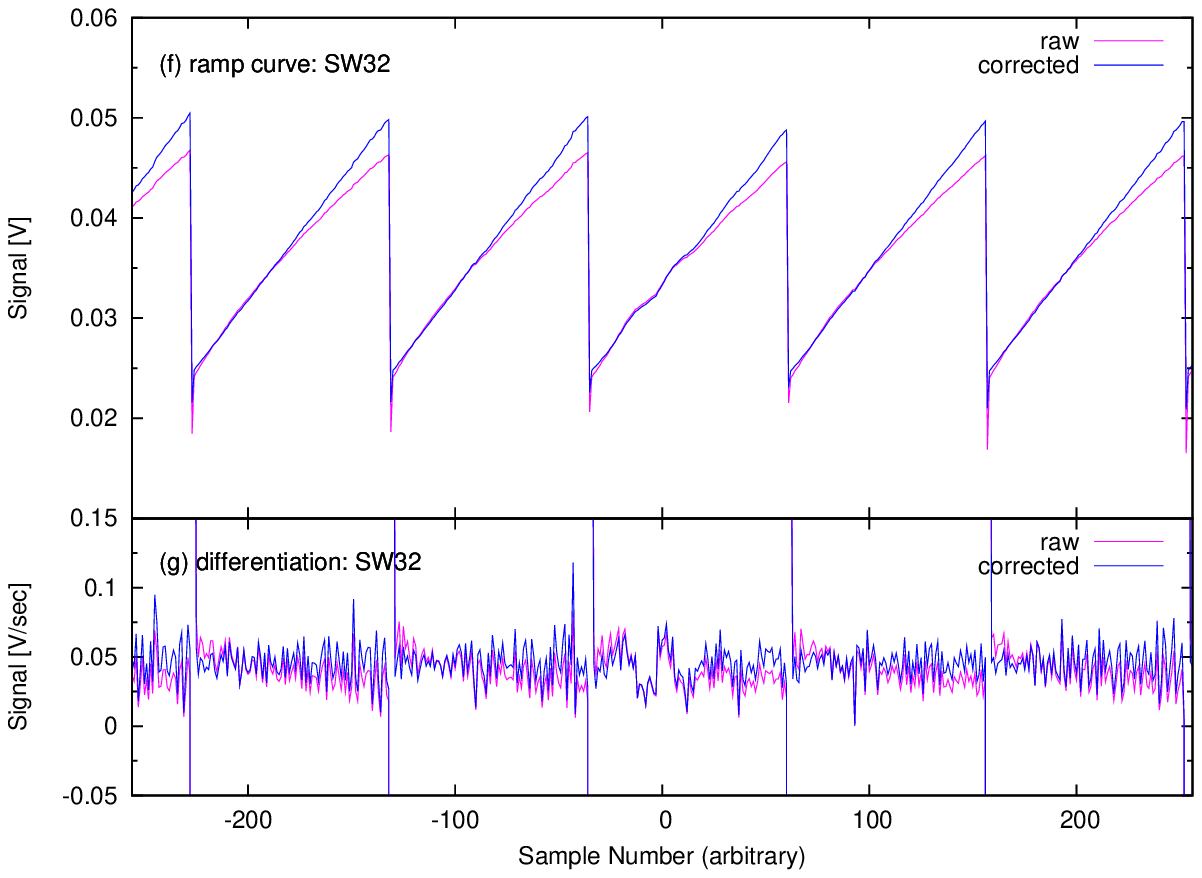} \\
	\FigureFile(80mm,60mm){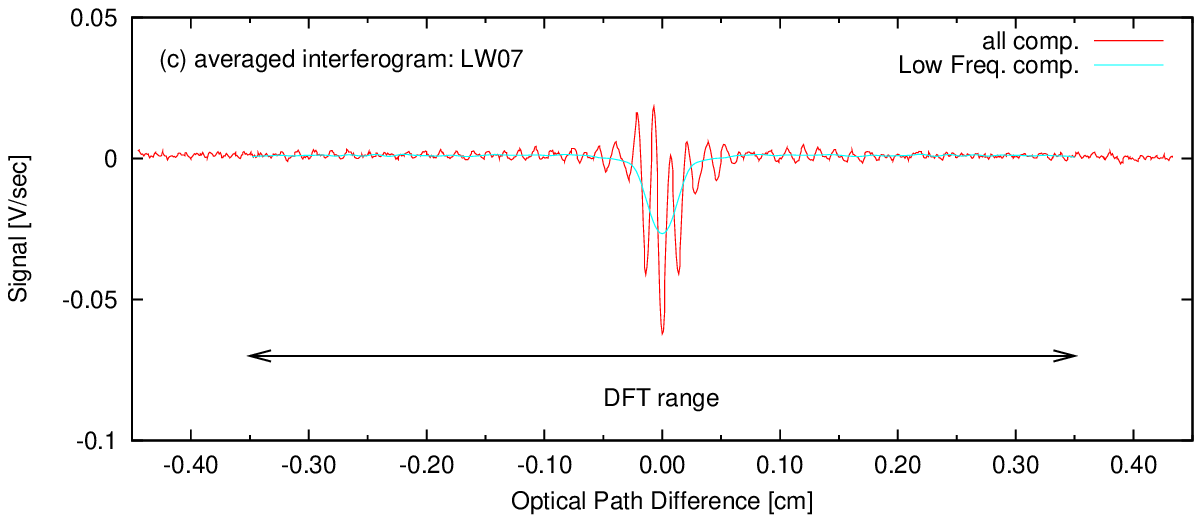} & \FigureFile(80mm,60mm){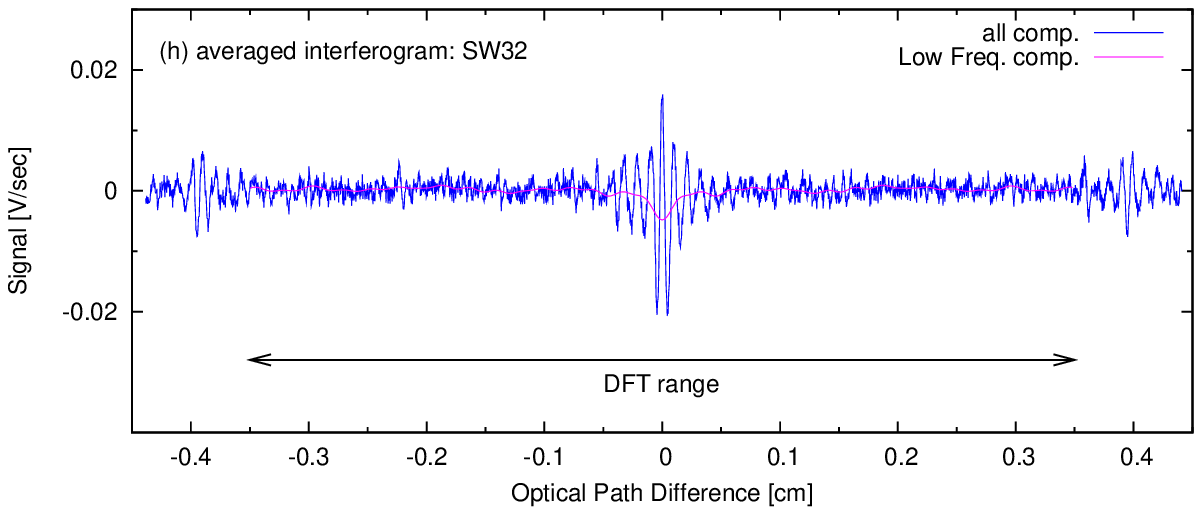} \\
	\FigureFile(80mm,60mm){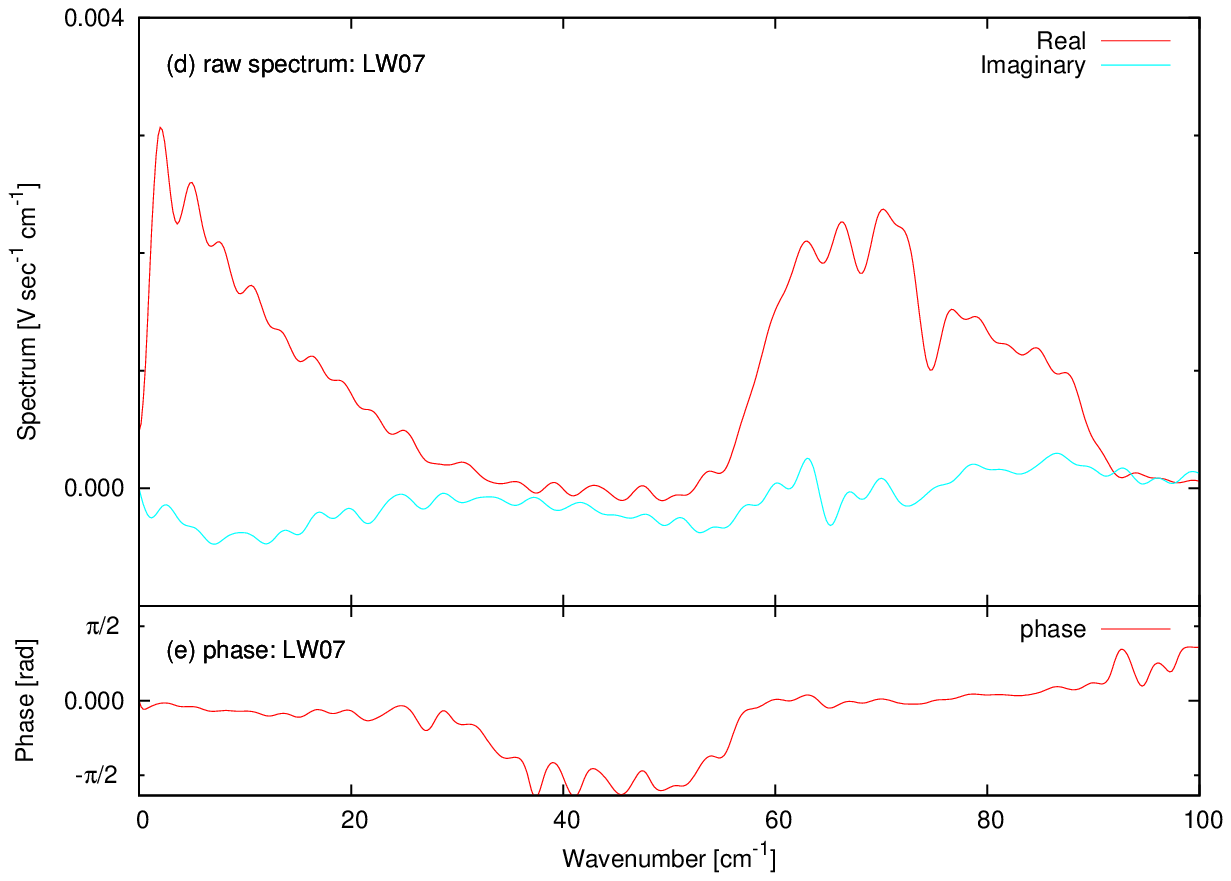} & \FigureFile(80mm,60mm){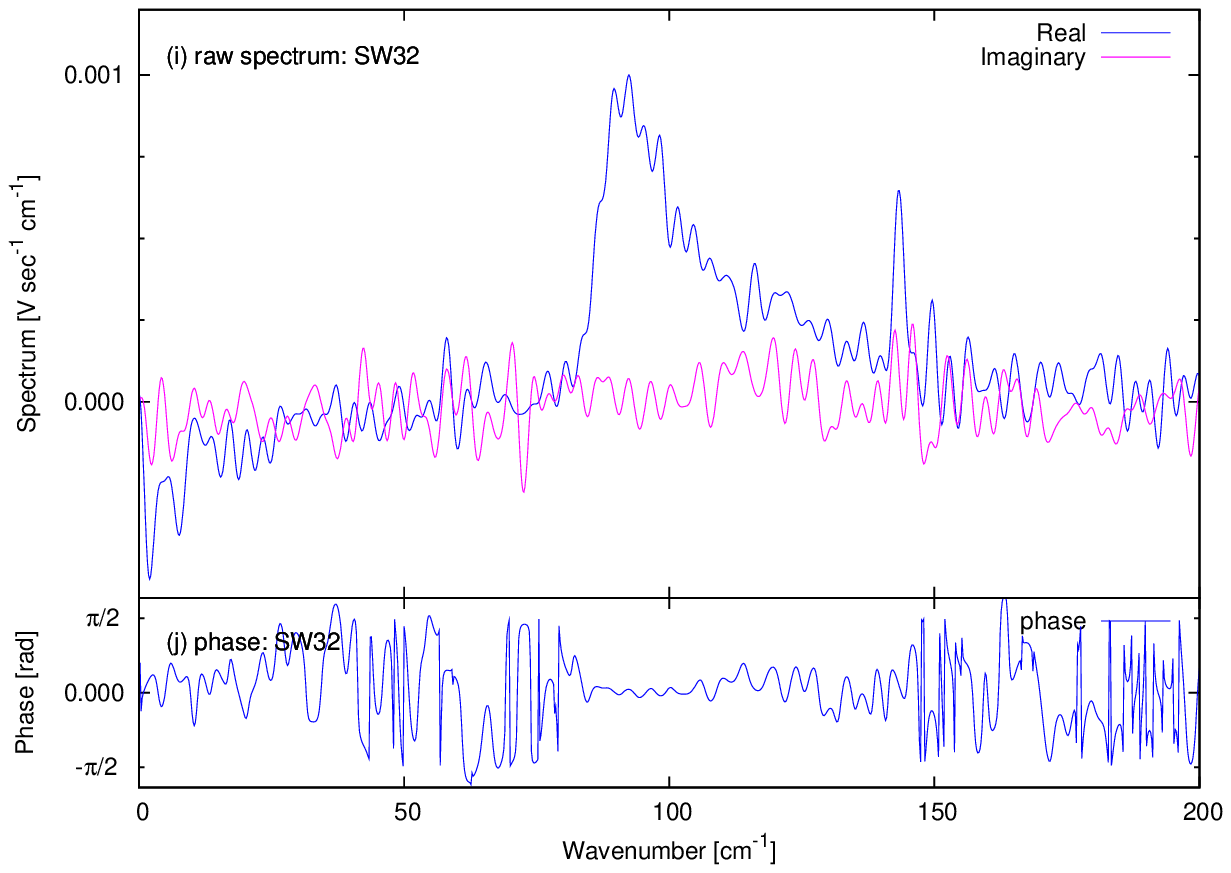} \\
    \end{tabular}
  \end{center}
  \caption{An Example of the FIS-FTS signals, interferograms and spectra as an 
	demonstration, which is measurement of the internal blackbody taken by the SED mode.  
	The left panels are plots of the LW detector and the right panels are of the SW 
	detector. Top panels (a) and (f) are raw signals of the LW07 and SW32 pixels, which 
	are typical pixels of each detector.  Panels of (b) and (g) show differentiations 
	of raw signal in time domain.  The corrected signals of the non-linearity related 
	to the readout electronics are also plotted on the raw signals.  The middle panels 
	(c) and (h) indicate the averaged interferograms of twelve scans (for same scan 
	direction).   The bottom panels (d) and (i) show raw spectra of the interferograms shown in 
	(c) and (h). Both real and imaginary part are plotted, and the phase signal are indicated 
	in (e) and (j).   Low frequency structures appeared in (d) and (i) come from the distortion 
	of the interferograms.   The contribution of low frequency components ($<$ 50 cm$^{-1}$) 
	are plotted on the interfergrams (c) and (h), which are derived by the inverse Fourier 
	transformation.}
  \label{fig:if_signal}
\end{figure*}

An example of the detector signal is shown in 
figures \ref{fig:if_signal}(a) and \ref{fig:if_signal}(f). 
The output signal is an integration ramp curve with periodic resets 
due to the CTIA readout circuit. By differentiating the signal, a 
nominal intereferogram including a DC component is derived 
(figures  \ref{fig:if_signal}(b) and \ref{fig:if_signal}(g)). 
The periodic feature in the reset interval as mentioned in 
section \ref{sss:detector} 
can be clearly seen. The periodic structure in the reset interval and 
random spikes due to cosmic ray hits must be removed. The standard 
data reduction flow for the FIS-FTS is described in \citet{Murakami2008}. 
A typical FIS-FTS interferogram after processing by the standard data 
reduction tool is shown in figures  \ref{fig:if_signal}(c) and 
\ref{fig:if_signal}(h). 
The left panels in figure  \ref{fig:if_signal} 
are plots of the LW detector (LW07) and the right panels are of the SW 
detector (SW32) using the internal blackbody source in the SED mode. 
This interferogram still has asymmetrical feature at the optical zero 
position, which can be clearly seen in the LW detector 
(figure  \ref{fig:if_signal}(c)). 
The distortion of the interferogram comes from the detector transient 
response (time is from left to right in the figure). Due to the 
difference in the time constant between upward and downward signals, 
the baseline is decreasing near the optical zero position. The 
influence of the distortion of interferograms is discussed in next 
section.

%\ref{fig:if_signal}(a),(b),(c),(d),(e),(f)

\subsubsection{Channel Fringes}

\begin{figure}
  \begin{center}
%    \begin{tabular}{c}
	\FigureFile(80mm,50mm){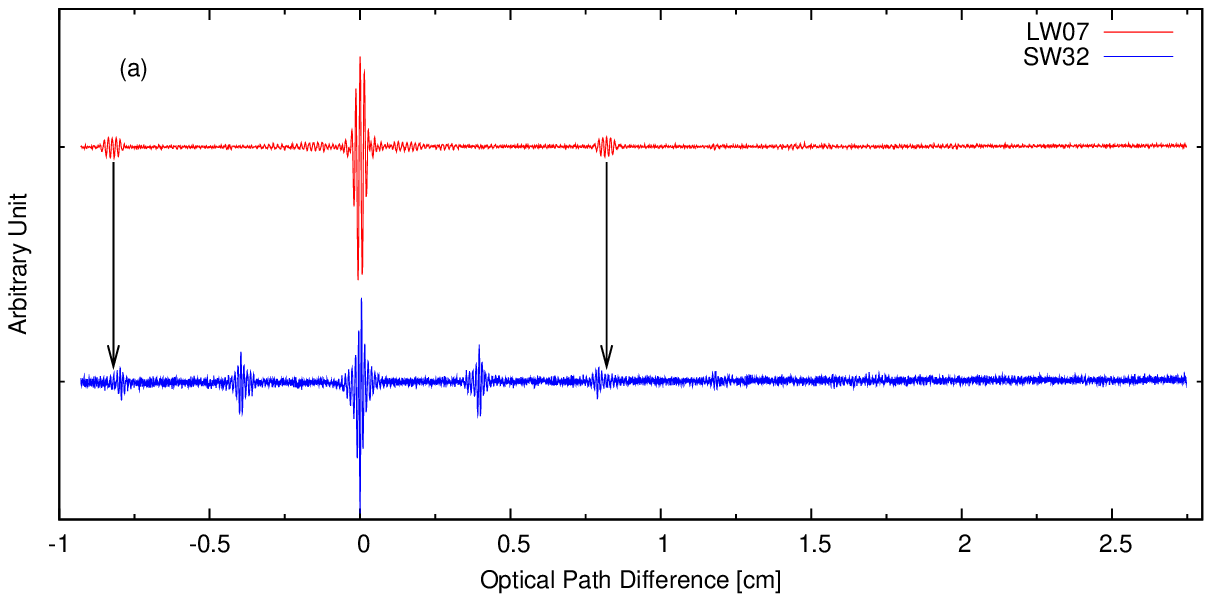} \\
	\FigureFile(80mm,60mm){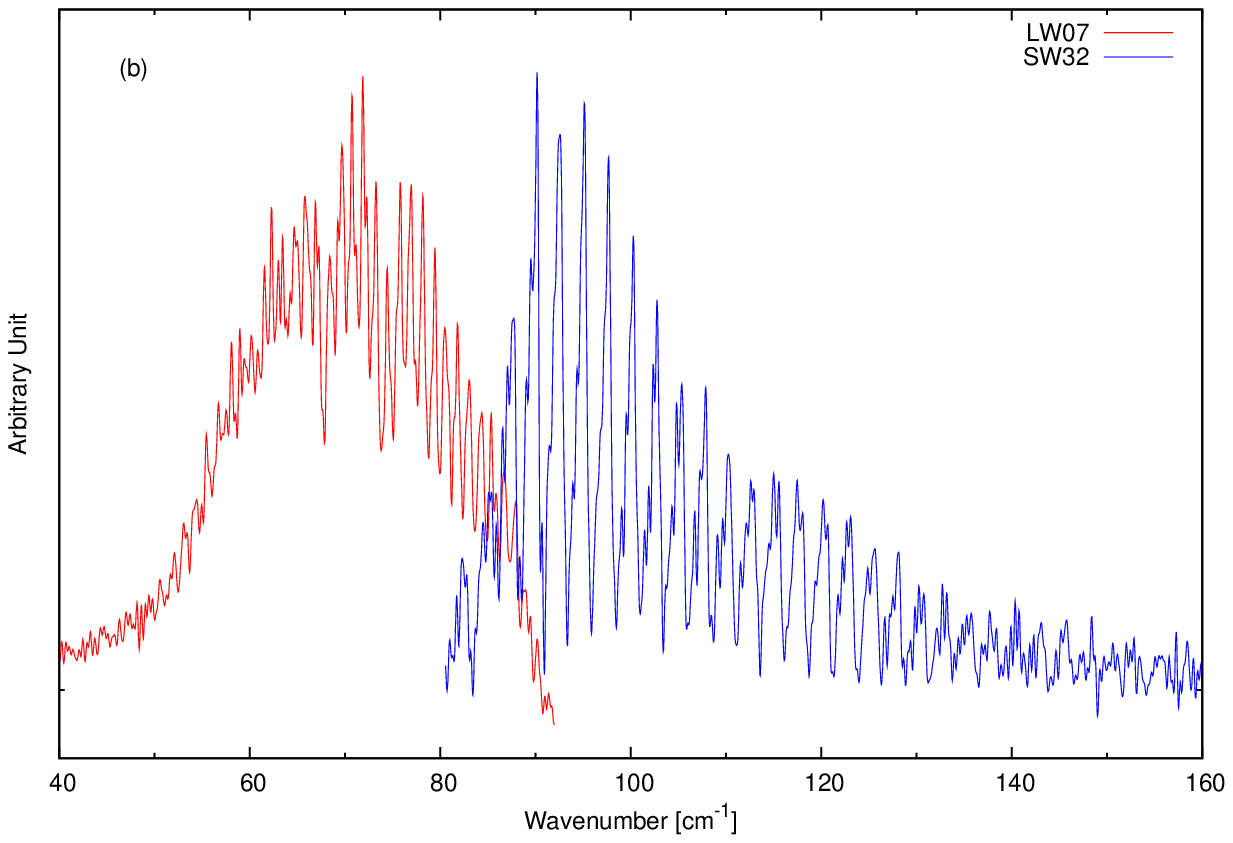}
%    \end{tabular}
  \end{center}
  \caption{An example of spectra taken by the full res. mode, which are spectra of the 
	telescope lid.  The top panel (a) shows interferograms averaging 6 scans of same 
	scanning direction.  There are sub-structures in the interferograms.  The upper
	interferogram is of LW07, and the lower one is of SW32.  In the SW32 interferogram, 
	same sub-feature seen in the LW07 can be recognized adding to the own
	periodic sub-structures.  The bottom panel (b) indicates the Fourier transformed 
	spectra of the interferograms in the top panel.  There are fringe structures in 
	both SW and LW spectra, whose periods are different from each other.}
  \label{fig:fringe}
\end{figure}

As shown in figure  \ref{fig:fringe}, 
there is a strong fringe structure in both SW and LW signals, especially 
in the SW. A periodic structure can be seen in the SW interferogram 
with about $\pm$0.4 cm 
periodicity, occurring on both sides of the optical zero position 
(figure  \ref{fig:fringe}(a)).  
This fringe can be identified to the third order at least, and comes 
from multi-beam interference in the detector substrate, which is made 
of Ge crystal with 0.5 mm thickness. The optical path difference due to 
one reflection is 
0.5 mm $\times$ 4 (refraction index of Ge) $\times$ 2 = 4 mm, 
and the actual value derived from the signal is 3.946 mm. The 
reflectivity estimated from the fringe contrast is about 0.4, which is 
consistent with reflectivity expected from the refractive index of Ge 
(n = 4).

Another type of channel fringe exists in the FIS-FTS, which can be 
clearly seen in the LW interferograms shown in figure  \ref{fig:fringe}(a). 
The first order of peaks appear on both sides of the main peak with 
about 0.8 cm separation in optical path difference.  The corresponding 
fringe pitch in the spectra is about 1.21 cm$^{-1}$ 
(figure  \ref{fig:fringe}(b)), 
with modulation amplitude significantly smaller than the other fringe 
seen in the SW spectrum. The observed fringes are nearly reproduced by 
a two parallel multi-reflection model with 4.1 mm gap and with about 20\% 
reflectivity. A plausible explanation of this phenomena is the two 
blocking filters put on the optical path just after the collimator 
mirror (BF\#1 \& \#2 in figure \ref{fig:FIS_optics}), with a physical 
gap of about 4 mm. The fact that this type of 
fringe can be seen in only sky signals, and not in the internal 
blackbody source, supports the multi-reflection model due to the 
blocking filters. The reflectivity of 20\%, 
however, is rather high compared to the laboratory measurements of the 
filters (about 10\%). 
In SW detector spectrum, this type of fringe is not clear because the 
other fringe dominates, but it can be found in the SW interferograms.

As the result of the fringe structure, the spectral response of the 
FIS-FTS changes rapidly with frequency. It is important to correct the 
fringe structure to derive line intensities. Details of the treatment 
of the fringe in the data reduction are described in \citet{Murakami2008}.
  
\subsubsection{Spectral Resolution}

The spectral resolution of the FIS-FTS is evaluated only in orbit 
using well known far-infrared lines of several sources. The expected 
spectral resolution from the maximum optical path difference 
($L \sim$ 2.7 cm) 
is about 1/2$L \sim$ 0.19 cm$^{-1}$ 
(rectangle apodization) for the full-res. mode. Some fine structure 
lines, like [CII] (63.40 cm$^{-1}$), 
[NII] (82.10 cm$^{-1}$), 
[OIII] (113.18 cm$^{-1}$), 
are detected from many observations. An example of the detected lines 
is shown in figure  10 in \citet{Kawada2007}, 
which is the [CII] line of the M82 galaxy. Physical line broadening of 
sources are negligible for the FIS-FTS because of its rather low 
spectral resolution ($R \sim$ 300 at [CII] line).
A spectral resolution derived from the observed lines is in good 
agreement with the expected value, which means the mirror driving 
mechanism works very well over the full travel span.  There is no 
evidence of wavenumber and pixel position dependence to the spectral 
resolution. A wavenumber scale correction, however, is required for 
each pixel according to the actual optical path length, which is 
longer than scale length because of the off-axis beam.  The maximum 
scale factor is expected to be less than 0.2\%, 
for ideal, which is consistent with the observed trend.

%\ref{fig:fringe}

\subsection{Stability and Reproducibility}

There are two types of internal calibration sources in the FIS-FTS: one 
is a micro lamp put in front of the detectors, the other is a small 
blackbody source put at the opposite port of the input polarizer. In 
the AOT of the FIS-FTS, the micro lamps are flashed at the beginning 
and end of each observation. The duration of each observation is about 
30 minutes. The blackbody source is activated prior to observing the 
sky for about 7 minutes, with temperature fixed to about 38 K. Since 
the blackbody source is quite stable and reproducible, it can be used 
as the health check of the FIS-FTS.

In the one and half year lifetime of the AKARI cryogens, the FIS-FTS 
has been operated for about 600 sets of pointed observations, 
corresponding to a total 
operation time of the FIS-FTS of about 200 hours. Using all observation 
data, we monitored the spectra of the internal blackbody source. Except 
for a few observations, which may be affected by radiation effects, the 
spectra of the internal blackbody source taken by the FIS-FTS are well 
reproducible within about $\pm$ 10\% 
variation. If the spectra are scaled with the integrated power of 
spectra, the variation of the spectra is reduced to less than $\pm$ 5\% 
for whole effective wavenumber region. The FIS-FTS system is quite 
stable and reproducible in spectra over the whole observation period. 

The FIS-FTS covers a wide wavenumber region with two kinds of detector 
arrays. To get the whole spectrum, we must combine two spectra taken by 
both detector arrays. The effective wavenumber overlap of the two 
detector arrays is around 85 -- 90 cm $^{-1}$, 
which is helpful to check the reproducibility of the spectra for both 
detector arrays.

\section{Discussions}

The overall spectral response of the FIS-FTS is determined by three 
factors, mainly; the filter transmission including multi-beam 
interference, the modulation efficiency due to the optical alignment, 
and the detector spectral response including the multi-beam reflections. 
In addition to the optical response, the transient response of the 
photoconductive detectors affects the resulting spectra through the 
distortion of the interferograms. In this section, we discuss about the 
performance as the FTS with photoconductive detectors. 

The fringe structure in the spectra complicates derivation of line 
intensities. Though the multiple reflections in the substrate of the 
SW detector contribute to the improvement of the total responsivity, 
the mulit-beam interference fringes make it difficult to measure the 
spectra. The fringes could be modeled by multi-beam interference of 
the detector substrate or two parallel blocking filters. The actual 
fringe shape is, however, different from an ideal Airy function with 
proper parameters.  It may be effects of the convergent beam or 
imperfect parallelism of the two flat surfaces. In practice, correction 
using templates made from actual data are effective. The continuum 
spectra derived from the short range interferograms are equal to the 
smoothed spectra with high resolution. If the responsivity at the line 
center is higher than the smoothed responsivity, a line to continuum 
ratio should be an overestimation, and vice versa. The data analysis 
and calibration methods are discussed in detail by \citet{Murakami2008}.

The FIS-FTS is the first FTS operating with an extrinsic photoconductive 
detector in space.  As shown in figure  \ref{fig:if_signal}, 
interferograms are distorted by the detector transient response. As the 
result of the distorted interferogram, the spectrum has a prominent 
structure at lower frequency as shown in figure  \ref{fig:if_signal}(d)(i). 
The low frequency component comes from the bump in the baseline around 
the optical zero position. It can be clearly seen as the inverse 
Fourier transform of the low frequency component (see figure 
\ref{fig:if_signal}(c)(h)). For the nominal 
operation of the FIS-FTS, the effective wavenumber ranges correspond to 
the modulation frequency ranges of 3.5 - 7 Hz and 7 - 15 Hz for the LW 
detector and the SW detector, respectively. As shown in figure  \ref{fig:if_signal}(d)(i), 
which are the data of LW07 and SW32 pixels, the real spectrum may be 
isolated from the low frequency component, which comes from the detector 
transient response artifact. The influence of the detector transient 
effect on the spectra are different from pixel to pixel. The LW 
detector has an especially wide variation in the transient response 
from pixel to pixel. Therefore, the distortion of the real spectrum 
by the detector transient effect should be evaluated carefully for 
precise analysis. In the zero order analysis, however, it may be 
acceptable to ignore the detector transient effects. The calibration 
strategy and accuracy of the FIS-FTS are discussed in \citet{Murakami2008}.

We can derive the interferometric efficiency by comparing the integrated 
power of the spectrum in the effective wavenumber range with the incident 
power. The overall efficiencies are around 50\% and 20\% for SW and 
LW detectors, respectively (open squares in figure \ref{fig:flat}). 
Considering the optical modulation efficiency 
(see \ref{sss:modulation_efficiency}), 
the degradation factors due to the detector transient response are 
$\sim$0.7 and $\sim$0.25 for SW and LW detectors, respectively.  
In a future paper, we will 
discuss the correction of the interferograms in the time domain 
considering the physical model of the photoconductive detector response. 
The operation of the FIS-FTS, unfortunately, was not optimized for 
photoconductive detectors because of our poor experience of FTS with 
photoconductive detectors. For future missions, such as SPICA which is 
the large infrared telescope mission \citep{Nakagawa07}, 
a FTS with photoconductive detectors is one of the possible choices for 
a high performance spectrometer. To bring out the maximum performance, 
we have to study the FTS with photoconductive detectors and a proper 
physical model of the photoconductive detector. The data produced by 
the FIS-FTS are quite valuable for this purpose as a precursor mission.

\section{Summary}

The FIS-FTS instrument is the first imaging Fourier transform 
spectrometer for far-infrared astronomy operated in space. The FIS-FTS 
was installed as an function of the Far-Infrared Surveyor (FIS) on the 
AKARI satellite, which was launched on February 21, 2006 (UT).  All 
functions of the FIS-FTS worked in orbit as well as they did in the 
laboratory. The performance of the FIS-FTS also achieved the same level 
seen in the laboratory. During the one and half year lifetime of the 
FIS instrument, the total operation time reached more than 200 hours or 
about 600 sets of pointed observations. The sensitivity, spectral response and 
resolution are quite stable for all observations.   

A unique feature of the FIS-FTS is the detector system; it is the first 
case to adopt a photoconductive detector array to FTSs.  Photoconductive 
detectors have high responsivity on one side, but severe transient 
response effect on the other side. As the result of this combination, 
the interferograms taken by the FIS-FTS are distorted and the spectra 
are affected by the transient response of the detectors. These 
difficulties in measuring the spectra should be avoided or reduced in 
future missions, if adopting a FTS with photoconductive detectors. 
The FIS-FTS provides good data for studying the detector transient 
effects on FTSs.

\bigskip     
 
The AKARI project, previously named ASTRO-F, is managed and operated by 
the Institute of Space and Astronautical Science (ISAS) of Japan 
Aerospace Exploration Agency (JAXA) in collaboration with groups in 
universities and research institutes in Japan, the European Space 
Agency, and Korea. We thank all the members of the AKARI/ASTRO-F 
project for their continuous help and support. FIS was developed in 
collaboration with Nagoya University, ISAS, University of Tokyo, 
National Institute of Information and Communications Technology (NICT), 
National Astronomical Observatory of Japan (NAOJ), and other research 
institutes.  Optical design of the FIS-FTS was done with the help of 
JASCO, and mechanical design was done with the Technical Center of 
Nagoya University. QMC Instruments Ltd developed optical filters for 
the FIS instrument. Many graduate students, especially Mr. Yasushi 
Tsuzuku, Mr. Yoshikazu Kuwata, Mr. Hiroshi Utsuno, Mr. Keita Ozawa and 
Mr. Tetsuo Imamura, worked for development, fabrication and testing of 
the FIS-FTS. We thank all the members related to the FIS-FTS for their 
intensive efforts. We really appreciate the useful advice on FTSs 
from Prof. Donald E. Jennings at NASA/Goddard Space Flight Center.
We would like to thank Prof. Annie Zavagno and Dr. Jean-Paul Baluteau 
at the Laboratoire d'Astrophysique de Marseille in France, Peter Davis 
at Blue Sky Spectroscopy in Canada, and Prof. David Naylor and Brad Gom 
at the University of Lethbridge in Canada for their contributions to 
the FIS-FTS data reduction and valuable discussions.

%%%%%%%%%%%%%%%%%%%%%%%%%%%%%%%%%%%%%%%%%%%%%%%%

\newpage

%%%% Tables %%%%%
%
%%%
%
%
%================= figures ===================
%\begin{figure}
%  \begin{center}
%    \FigureFile(80mm,80mm){fig1.eps}
%    %%% \FigureFile(width,height){filename}
%  \end{center}
%  \caption{This is the first figure.}\label{fig:sample}
%\end{figure}


\begin{thebibliography}{}
% Journals(e.g. A\&A,ApJ,AJ,NMRAS,PASP ...)
% Authors, Year, Journal, Vol#, Page#
% Journal Title Abbreviation >> http://www.asj.or.jp/pasj/Jabb.html
\bibitem[Doi et al. (2002)]{Doi02} 
	Doi, Y. et al. 2002, Adv. Space Res., 30, 2099--2104
\bibitem[Fujiwara et al. (2003)]{Fujiwara03}
	Fujiwara, M., Hirao, T., Kawada, M., Shibai, H., Matsuura, S., Kaneda, H., 
	Patrashin, M. A., \& Nakagawa, T. 2003, \ao, 42, 2166--2173
\bibitem[Hirao et al. (1996)]{Hirao96}
	Hirao, T., Matsumoto, T., Sato, S., Ganga, K., Lange, A. E., 
	Smith, B. J., \& Freund, M. 1996, \pasj, 48, L77--L82
\bibitem[Kaneda et al. (2002)]{Kaneda02}
	Kaneda, H., Okamura, Y., Nakagawa, T., \& Shibai, H. 2002, 
	Adv. Space Res., 30, 2105--2110
\bibitem[Kaneda et al. (2007)]{Kaneda2007}
	Kaneda, H., Kim, W., Onaka, T., Wada, T., Ita, Y., 
	Sakon, I., \& Takagi, T. 2007, \pasj, 59, 423--427
\bibitem[Kawada et al. (2007)]{Kawada2007}
	Kawada, M. et al. 2007, \pasj, 59, 389--400
\bibitem[Kessler et al. (1996)]{Kessler96}
	Kessler, M. F. et al. 1996, \aap, 315, L27--L31
\bibitem[Kessler (2002)]{Kessler02}
	Kessler, M. F. 2002, Adv. Space Res., 30, 1957--1965
\bibitem[Kunde et al. (1996)]{Kunde96}
	Kunde, V. G. et al. 1996, Proc. SPIE, 2803, Ed. L. Horn, 162--177
\bibitem[Martin \& Puplett (1969)]{Martin69}
	Martin, D. H., \& Puplett, E. 1969, Infrared Phys., 10, 105 
\bibitem[Mather (1993)]{Mather93a}
	Mather, J. C. 1993, Proc. SPIE, 2019, Ed. M. S. Scholl, 146--157
\bibitem[Mather et al. (1993)]{Mather93b}
	Mather, J. C., Fixsen, D. J., \& Shafer, R. A. 1993, 
	Proc. SPIE, 2019, Ed. M. S. Scholl, 168--179
\bibitem[Murakami et al. (1996)]{Murakami96}
	Murakami, H. et al. 1996, \pasj, 48, L41--L46
\bibitem[Murakami et al. (2007)]{Murakami2007}
	Murakami, H. et al. 2007, \pasj, 59, 369--376
\bibitem[Murakami et al. (2008)]{Murakami2008}
	Murakami, N. et al. 2008, \pasj, XX, XXX--XXX
\bibitem[Nagata et al. (2004)]{Nagata04} 
	Nagata H., Shibai H., Hirao T., Watabe T., Noda M., Hibi Y., 
	Kawada M., \& Nakagawa T. 2004, IEEE Trans. Elec. Devices, 51, 270 
\bibitem[Nakagawa et al. (2007)]{Nakagawa2007}
	Nakagawa, T., et al. 2007, \pasj, 59, 377--387
\bibitem[Nakagawa \& Murakami (2007)]{Nakagawa07}
	Nakagawa, T., \& Murakami, H. 2007, Adv. Space Res., 40, 679--683
\bibitem[Neugebauer et al. (1984)]{Neugebauer84}	
	Neugebauer, G. et al. 1984, \apj, 278, L1--L6
\bibitem[Ohyama et al. (2007)]{Ohyama2007}
	Ohyama, Y. et al. 2007, \pasj, 59, 411--422
\bibitem[Okada et al. (2008)]{Okada2008}
	Okada, Y. et al. 2008, in preparation %\pasj, XX, XXX--XXX
\bibitem[Onaka et al. (1996)]{Onaka96}
	Onaka, T., Yamamura, I., Tanabe, T., Roellig, T. L., \& Yuen, L. 1996, 
	\pasj, 48, L59--L63
\bibitem[Onaka et al. (2007)]{Onaka2007}
	Onaka, T. et al. 2007, \pasj, 59, 401--410
\bibitem[Shibai et al. (1996)]{Shibai96}
	Shibai, H., Okuda, H., Nakagawa, T., Makiuti, S., Matsuhara, H., 
	Hiromoto, N., \& Okumura, K. 1996, \pasj, 48, L127--L131
\bibitem[Shirahata et al. (2004)]{Shirahata04}
	Shirahata, M. et al. 2004, Proc. SPIE, 5487, Ed. J. C. Mather, 369--380
\bibitem[Takahashi et al. (2003)]{Takahashi03}
	Takahashi, H., Kawada, M., Murakami, N., Ozawa, K., 
	Shibai, H., \& Nakagawa, T. 2003, 
	Proc. SPIE, 4850, Ed. J. C. Mather, 191--201
\bibitem[Takahashi et al. (2008)]{Takahashi2008}
	Takahashi, H. et al. 2008, in preparation %\pasj, XX, XXX--XXX
\bibitem[Tanaka et al. (1996)]{Tanaka96}
	Tanaka, M., Matsumoto, T., Murakami, H., Kawada, M., 
	Noda, M., \& Matsuura, S. 1996, \pasj, 48, L53--L57
\bibitem[Werner et al. (2004)]{Werner04}
	Werner, M. W. et al. 2004, \apjs, 154, 1--9
\bibitem[Yasuda et al. (2008)]{Yasuda2008}
	Yasuda, A. et al. 2008, in preparation %\pasj, XX, XXX--XXX

%\bibitem[Aauthor et al.(2001)]{key-1}
%   Aauthor, A., Bauthor, B., Cauthor, C.\ 2001, PASJ, vol, page
% Books
%\bibitem[Aauthor \& Author(2001a)]{key-2}
%   Aauthor, A., Author, B.\ 2001, Name of Book(Publisher, Tokyo) ch0
% Books
%\bibitem[Aauthor \& Bauthor(2001b)]{key-3}
%  Aauthor, A., Bauthor, B.\ 2001, Name of Book(Publisher, Tokyo) page0
%......
% Editorial Books
%\bibitem[Dauthor(2001)]{key-n}
%  Dauthor A. A.\ 2001, in Name of Book,
%   ed Editor D.\ Editor(Publisher, Tokyo) page0
\end{thebibliography}
\end{document}